# Lee, Han, and Kim Reply:

In Letter [1], we reported our finding on the physical origin of wide band-gap opening in planar nanostrips, as derived by application of periodic width modulations in the magnetic waveguides. K. Di *et al*. in their Comment[2], however, argued that the band gap can be reduced remarkably by applying a linear combination of symmetric and antisymmetric fields (see 'A+B' in the inset of Fig. 1(a)). They also insisted that they could find a complete set of magnonic bands based on all of the mode symmetries. However, their work does not constitute grounds for judging our method "wrong" and our conclusions "erroneous". Moreover, the excitation field alluded to in the Comment is not sufficiently general to obtain the complete set of magnonic band structures, but rather remains specific. The grounds of these conclusions, drawn from our further work, are the following.

In Fig.1(b), the superposition of the two blue and red dispersion curves obtained independently from symmetric ('B') and antisymmetric ('A') excitation fields, respectively, agrees with the dispersion (green) curves obtained from the superposition of those two excitation fields ('A+B'). This indicates that the modes excited independently from the symmetric and antisymmetric fields are not coupled. Thus, the change of the position and width of the band gaps in the dispersion curves for consideration of the 'A+B' field is due only to the existence of addition modes driven by the asymmetric field [2]. Consequently, our results driven by the symmetric field ('B') are physically correct. In Fig.1(b), the bands of the antisymmetric modes are located in between the bands of the symmetric modes, which effects a significant band-gap decrease. However, we newly found that, even with respect to both the symmetric and antisymmetric modes, wide band gaps on the order of 10 GHz can be produced simply by changing the periodicity of the width-modulated wave guides, as shown in Fig. 1(c). Our proposed magnonic crystals still offer wide band gaps on the order of 10

GHz and they are applicable to spin-wave filters [3.4].

K. Di *et al.*[2] also stated that to excite full band structures, they used the 'A+B' magnetic field. However, this method is not generalizable either. Figure 2(a) shows the dispersion curves for spin waves in a *single-width* nanostrip from the 'A', 'B', and 'A+B' excitation fields. The symmetric field excites the first lowest ($m$=1) mode, but the antisymmetric field the second lowest ($m$=2) mode. The superposition of the two fields excites the two lowest modes, as seen in Fig. 2(b). Moreover, a very specific field exerted only at one spot position (see Ref.[5]) excites an additional third width mode ($m$=3) as well as both the two lowest modes (see Fig. 2(c)). Therefore, although the 'A+B' field used in the Comment[2] is not general to obtain the complete magnonic bands, consideration of symmetric and antisymmetric excitation fields and their superposition enables study of additional antisymmetric modes in width-modulated nanostrips.

Ki-Suk Lee,[1] Dong-Soo Han,[2] and Sang-Koog Kim[2]*

1. School of Mechanical and Advanced Materials Engineering, UNIST, Ulsan 689-798, Republic of Korea

2. National Creative Research Initiative Center for Spin Dynamics & Spin-Wave Devices and Nanospinics Laboratory, Research Institute of Advanced Materials, Department of Materials Science and Engineering, Seoul National University, Seoul 151-744, Republic of Korea

*Corresponding author. sangkoog@snu.ac.kr

[5] The spot is 1.5x1.5x10nm$^3$ in size and positioned at $y = 19.5$ nm (black). This off-centered spot position is chosen to avoid excitation of only the symmetric width-modes for the case where the symmetric field is applied at the center position of $y = 0$ nm.

FIG.1. (a) Geometry and dimensions of width-modulated nanostrip. A sinc field pulse was applied to the 1.5x30x10 nm$^3$ central area (dark blue). The inset shows four different excitation fields as indicated. (b), (c) Dispersion curves of spinwave modes in width-modulated strips of indicated values of [$P_1$, $P_2$], as obtained from FFTs of temporal $M_z/M_s$ oscillations along the $x$ axis at position $y$=19.5 nm to observe both the symmetric and antisymmetirc width-modes.

FIG.2. Dispersion curves for 30 nm single-width nanostrip excited by (a) 'A', 'B', (b) 'A+B', and (c) 'Spot' fields.

FIG. 1.

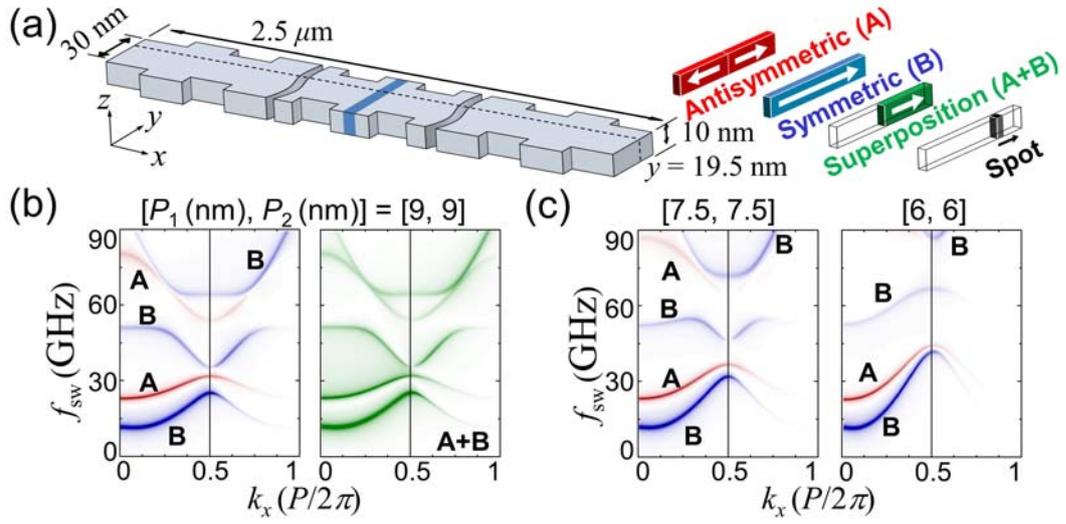

FIG. 2

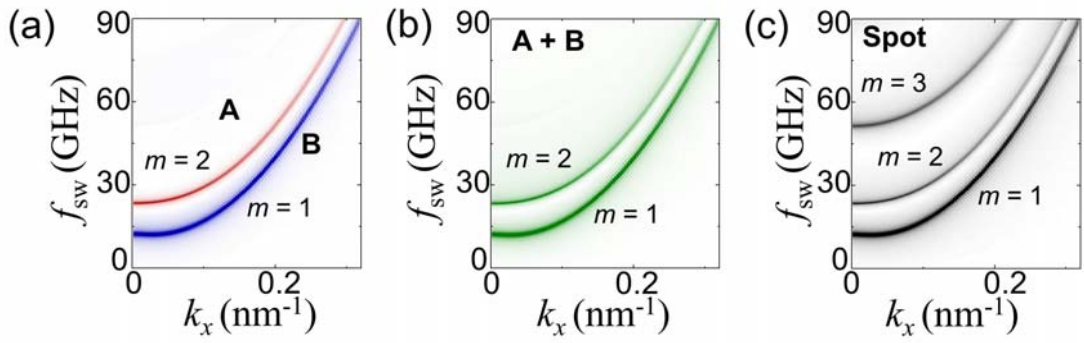